\documentclass[useAMS,usenatbib,draftcopy]{mn2e}

\usepackage{graphicx}
\usepackage{graphics}
\usepackage{rotating}
\usepackage[light]{}

\newcommand{\mnras}{MNRAS}
\newcommand{\apj}{ApJ}
\newcommand{\apjl}{ApJL}
\newcommand{\aap}{AAP}
\newcommand{\apjs}{ApJS}
\newcommand{\aj}{AJ}
\newcommand{\bain}{BAIN}
\newcommand{\nat}{Nat}

\def\kms      {\ifmmode{\rm km\,s}^{-1} \else km\,s$^{-1}$\fi}
\def\mujybm{\ifmmode{\rm \mu Jy}\,{\rm beam}^{-1}\else${\rm \mu}$Jy\,beam$^{-1}$\fi}
\def\ltsim{\ifmmode\stackrel{<}{_{\sim}}\else$\stackrel{<}{_{\sim}}$\fi}
\def\gtsim{\ifmmode\stackrel{>}{_{\sim}}\else$\stackrel{>}{_{\sim}}$\fi}
\def\S4195{41.95+575}
\def\S4331{43.31+592}

\begin{document}   
\title[Radio source calibration at 30 GHz]{Radio source calibration for the VSA and other CMB instruments at around $30$~GHz}

\author[Hafez {\it et al.}]{Yaser\,A.\,Hafez,$\!\!^{1,2}$\thanks{yhafez@kacst.edu.sa} Rod\,D.\,Davies,$\!^1$\thanks{rdd@jb.man.ac.uk} Richard\,J.\,Davis,$\!^1$\thanks{rjd@jb.man.ac.uk} Clive\,Dickinson,$\!^{3}$ 
\newauthor Elia\,S.\,Battistelli,$\!^{4,5}$ Francisco\,Blanco,$\!^1$ Kieran\,Cleary,$\!^6$ Thomas Franzen,$\!^7$
\newauthor Ricardo\,Genova-Santos,$\!^4$ Keith\,Grainge,$\!^7$ Michael\,P.\,Hobson,$\!^7$ Michael\,E.\,Jones,$\!^8$
\newauthor Katy\,Lancaster,$\!^9$ Anthony\,N.\,Lasenby,$\!^7$ Carmen\,P.\,Padilla-Torres,$\!^4$ 
\newauthor Jos\'e\,Alberto\,Rubi\~{n}o-Martin,$\!^4$ Rafael\,Rebolo,$\!^4$ Richard\,D.\,E.\,Saunders,$\!^7$
\newauthor Paul\,F.\,Scott,$\!^7$ Angela\,C.\,Taylor,$\!^{8}$ David Titterington,$\!^7$ Marco\,Tucci,$\!^4$ 
\newauthor \& Robert\,A.\,Watson\,$\!^1$ \\
$^1$Jodrell Bank Observatory, The University of Manchester, Macclesfield, Cheshire, SK11~9DL, UK\\
$^2$National Center for Mathematics and Physics, KAST, PO Box 6086, Riyadh 11442, Saudi Arabia \\
$^3$Infrared Processing and Analysis Center, California Institute of Technology, M/S 220-6, 1200 E. California Blvd., Pasadena, 91125, U.S.A. \\
$^4$Instituto de Astrofis\'{\i}ca de Canarias, 38200 La Laguna, Tenerife, Canary Islands, Spain \\ 
$^5$Department of Physics \& Astronomy, University of British Columbia, 6224 Agricultural Road, Vancouver, British Columbia, V6T, 1Z1, Canada \\
$^6$California Institute of Technology, M/S 105-24, 1200 E. California Blvd., Pasadena, 91125, U.S.A. \\
$^7$Astrophysics Group, Cavendish Laboratory, University of Cambridge, Madingley Road, Cambridge, CB3 0HE, UK \\ 
$^8$Astrophysics Group, Denys Wilkinson Building, University of Oxford, Keeble Road, Oxford, OX1 3RH, UK \\
$^9$H H Wills Physics Laboratory, University of Bristol, Tyndall Avenue, Bristol, BS8 1TL, UK }

\date{2008-??-??}
\pagerange{\pageref{firstpage}--\pageref{lastpage}} \pubyear{2008}

\maketitle
\label{firstpage}

\begin{abstract} {
Accurate calibration of data is essential for the current generation
of  CMB experiments.  Using data from the Very Small Array (VSA), we
describe procedures which will lead to an accuracy of~1~percent or
better for experiments such as the VSA and CBI. Particular  attention is paid to the stability of the
receiver systems, the quality of the site and frequent observations of
reference sources.  At 30~GHz the careful correction for atmospheric
emission and absorption is shown to be essential for achieving 1~percent
precision.

The sources for which a 1 percent relative flux density calibration
was  achieved included Cas~A, Cyg~A, Tau~A and NGC7027 and the
planets Venus,  Jupiter  and Saturn.  A flux density, or brightness
temperature in the case of the planets, was derived at 33 GHz relative
to Jupiter which was adopted as  the fundamental calibrator. A
spectral index at $\sim30$ GHz is given for each.

Cas~A,Tau~A, NGC7027 and Venus were examined for variability. Cas~A was found to be decreasing at $0.394 \pm 0.019$ percent per year
over  the period March 2001 to August 2004.  In the same period Tau~A
was decreasing at $0.22\pm 0.07$ percent per year. A survey of the
published data showed that the planetary nebula NGC7027 decreased at
$0.16\pm 0.04$ percent per year over the period 1967 to 2003. Venus
showed an insignificant ($1.5 \pm 1.3$ percent) variation with
Venusian illumination. The integrated polarization of Tau~A at 33 GHz
was found to be $7.8\pm 0.6$ percent at pa $ = 148^\circ \pm 3^\circ$.}

\end{abstract}

\section{Introduction} 

With  the increasing sensitivity of  CMB experiments,  it is important
to have accurate calibration of the intensity  or temperature scale of
each experiment.  The cosmological significance  of the data is
directly dependent on this scale.  For example, the amplitude  of the
first peak in the CMB power spectrum is proportional to $\Omega_{\rm b}$ while the ratio
of the first to the third peak, derived from a mix of experiments
probing different angular modes $\ell$, is also sensitive to
$\Omega_{b}$.   Furthermore, the fractional uncertainty in the power spectrum is twice that of the fractional uncertainty in the temperature scale. Current CMB experiments would greatly benefit from a
coherent calibration  across experiments such as {\it Planck}, WMAP, CBI,
VSA etc. at an accuracy of  $1$\% or better.  The present
investigation provides a set of source  intensities calibrated to this
precision at a frequency of 33~GHz.

We have chosen the brightest planets and strongest radio sources to
calibrate the VSA on a frequent (typically few hours) basis during CMB observations.  The
basic  calibration of all the targets used in the VSA is in terms of
an assumed  brightness temperature of Jupiter (Watson et al.~2003;
Dickinson et al.~2004), which in turn, is tied to the CMB dipole (Hinshaw et al.~2008).

Inevitably the strongest sources used for the frequent calibration  of
VSA data are sufficiently extended (a few arcmin) that their total
flux densities require correction when applied to the $\sim$10 arcmin
resolution of the VSA at an accuracy of better than 1\%.  Cas~A, Tau~A
and  Cyg~A fall into this category although for Cyg~A the effect is
$\leq$0.3$\%$. The planets do not require such a  correction.

The polarization of the calibrators also needs to be taken  into
account. The most strongly polarized source is Tau~A.  The  integrated
emission from HII regions and the planets are not expected to be
polarized.  The radial polarization of the planets would only be of
concern if there were a significant phase effect in the brightness
distribution across the planet.

Time variability is an important consideration when using calibrators.
Many of the stronger extragalactic sources such as quasars have to be
rejected for this reason.  This is an integral part of the present
study.

The spectral index of a calibrator needs to be specified if the
calibration process is to be useful between experiments which have
different central frequencies and bandwidths even within the 1 cm
wavelength band investigated here.  In general, spectral index data
are   taken from the literature.

The main challenge in accurate calibration of  ground-based
observations at centimetric and millimetric frequencies is  provided
by the atmosphere.  A high dry site is a prerequisite.  The Iza\~na
observatory of the IAC has a proven record of observations in the
range  10-33 GHz (Davies et al.~1996).  An assessment of its properties at these
frequencies will be  given in a separate paper (Davies et al., in
prep.).

This paper is arranged as  follows.  Section 2 describes the relevant
features of the VSA used in  this investigation.  The philosophy of
the approach, including the method  of determining the atmospheric
corrections, are given in Section 3.  The  results given as flux
densities, or brightness temperatures in the case of  the planets, are
presented along with the adopted spectral indices in  Section 4, which
also includes a discussion of source variability and
polarization. The conclusions of this work are summarised in Section
5.

\section{The VSA and its calibration system} 

The advantage of using an  interferometer for ``point source'' flux
density measurements is that  the extended emission from the sky,
atmosphere and the ground can be  largely eliminated.  The VSA is
described in detail by Watson et al.  (2003), Scott et al. (2003) and
Dickinson et al. (2004).  A summary of its  main features is now given.

\subsection{The overall system} 

The Very Small Array (VSA) is a 14-element  interferometer operating
in the Ka band (26-36 GHz).  It is located on the  high and dry site
at the Teide Observatory, Iza\~na, Tenerife at an altitude  of 2340m.
The antenna of each element consists of a conical corrugated  horn
feeding a paraboloidal mirror.  Each antenna and its associated
cryogenically cooled receiver system can be placed anywhere on a 4-m x
3-m  tip-tilt table located in a metal enclosure to minimise any
emission  contribution from the ground.  Due to the geometry of the
table and the  enclosure, the VSA declination is restricted to the
range $-5^{\circ}$ to  $+60^{\circ}$.  Tracking of a field is
accomplished by a combination of  table  tilt and rotation of each
mirror.

Since the beginning of CMB observations  in September 2000 the VSA has
been operated in both a compact and extended  configuration.  In the
compact configuration the mirror diameters were 143mm  giving a primary
beamwidth (FWHP) of 4$.\!\!^{\circ}$6.  The geometrical  arrangement
of the  antennas on the table gave interferometer baseline lengths
ranging from  0.20m to 1.23m allowing angular multipoles of l = 150 to
900 to be probed  at the observing frequency of 34.1 GHz.  The
resolution of the array was  $\sim~17$~arcmin depending on the field
declination and hour angle range  observed.

The extended array, which became operational in October 2001, used  an
antenna configuration which covered a baseline range of 0.6m to 2.5m
thereby extending the range of angular multipoles to l = 1500 and gave
a  synthesized beamwidth (FWHP) of $\sim~11$~arcmin.  The larger
mirror  diameters provided a factor of 1.6 improvement in filling
factor and an  overall factor of $\sim~3$~in temperature sensitivity
(see Table  \ref{vsaspec}). Since September 2005 the VSA has operated in a third configuration, the super-extended array, with 60~cm aperture mirrors (Genova-Santos et al.~2008).

 The  VSA has operated at either 33 or 34.1 GHz with an instantaneous
bandwidth  of  1.5 GHz.  With an average system temperature of some
$35$~K the VSA  achieves an overall instantaneous point source
sensitivity of  $\approx 36$~and $\approx 6$~Jy~s$\!^{1/2}$ in the compact
and extended configurations  respectively;  the corresponding temperature
sensitivities are 40 and 15~mK~s$^{1/2}$.  The VSA is linearly
polarized in the vertical direction and  is therefore sensitive to
any linear polarization in the calibration sources.

\subsection{Present analysis} 

An interferometer array like the VSA  requires an accurate
determination of the position of the antennas to produce  reliable relative
observations at the 1\% level aimed at in this paper.   A maximum likelihood
technique has been applied to the calibrator source observations in
order to constrain such telescope parameters as the antenna positions,
effective observing frequencies and correlator amplitudes and phase
shifts  (Maisinger et al. 2003).  The stability of the
VSA is such  that these calibration observations are required
only every few hours.

 The overall  gain of each antenna is monitored continuously by means
of a noise  injection system.  A modulated noise signal is injected
into each antenna  via a probe in the horn and is measured using
phase-sensitive detection  after the automatic gain control (AGC)
stage in the IF system.  The  relative contribution of the constant
noise source to the total output  power from each antenna varies
inversely with system temperature and thus  a correction can be made
to the overall flux calibration.  This system  allows account to be
taken of the gain and variations in total set noise due, for example, to
atmospheric emission. It potentially provides an
indication of weather conditions and is a primary indicator for
flagging  weather-affected data in CMB observations.  In the face of
the stringent  requirements here of better than 1\% calibration
accuracy, this approach  had to be modified as described in the next
section.

\begin{table*} \centering \caption{Specification of 
the VSA in the compact and extended configurations.} \label{vsaspec} 
\begin{tabular}{ccccccccccc} \hline\noalign{\smallskip}
 &Compact &Extended\\ \hline
Declination range &$-5$ to $+60$ &$-5$ to $+60$ \\
Number of antennas (baselines) & 14(91) &14(91)\\
Range of baseline lengths &0.20~m to 1.23~m &0.6~m to 2.5~m\\
Centre frequencies &34 GHz &33, 34 GHz\\
Bandwidth of observation &1.5 GHz &1.5 GHz\\
System temperature, T$_{\rm sys}$ &$\approx 35$~K &$\approx 35$~K\\
Mirror diameters &143~mm &332~mm\\
Primary beam (FWHM) &4.6 &2.0\\
Synthesized beam (FWHM) &$\approx 17$~arcmin &$\approx 11$~arcmin\\
Range of angular multipole (l) &150 to 900 &300 to 1500\\
Point source sensitivity &$\approx 36$ Jy~s$^{1/2}$ &$\approx 6$~Jy~s$^{1/2}$\\
Temperature sensitivity per beam &$\approx 40$~mKs$^{1/2}$ &$\approx 15$~mKs$^{1/2}$\\
Polarization &linear (vertical) &linear (vertical)\\
\noalign{\smallskip}\hline
\end{tabular}
\end{table*}

\section{The source calibration programme}

We consider here the methods used to measure relative flux densities to an 
accuracy of 1 percent or better at a frequencies around 30~GHz.

\subsection{Calibration sources for CMB observations}

The sources available for this study are those bright enough to be 
detected with good signal to noise in say $20$~minutes of observing time.  
This limits them to $\geq$ 10 Jy in the two configurations, which allows the following sources to be used as primary calibrators: Tau~A, Cas~A, Cyg~A, Jupiter, Saturn and Venus. Observations on these sources of 10-30 minutes are interleaved within CMB fields.  These observations provide the 
data archive on which the present investigation is based.  Additional 
long-track observations covering $\pm$3 hr in hour angle were used to 
quantify elevation-dependent atmospheric effects.

\subsection{Data reduction procedures} 

The basic data reduction of VSA data followed the same procedure as
described in early VSA work (e.g. Taylor et al.~2003; Dickinson et
al.~2004). One radio source (or the CMB) is  calibrated against
another (the `` calibrator '') by converting the units of the 91  raw visibilities
from correlator units to to flux densities via the primary `` calibrator '' for each baseline
and then  combining them to give a mean flux density for the source.
The resulting  91 baseline visibility tracks are checked by eye and
interactively  searched for residual contaminating signals.  Due to the large number of calibrator observations we were able to be fastidious about data quality. Some 20\%
of the data are  removed in this way in addition to the more obvious
errors due to hardware  failure and bad weather.  The elevation of
each source was recorded and  could be used to  make subsequent
corrections for atmospheric effects.
                                                                        
\subsection{Choice of fundamental reference source} 

The flux  calibration of VSA data is tied to observations of
Jupiter which is  widely used as a calibrator for CMB observations in
the range 10-500 GHz.   Our early VSA results (see e.g. Scott et al. 2003;
Taylor et al. 2003) were based  on a brightness temperature T$_{J}$ =
152$\pm$5~K at 32~GHz, extrapolated to 33/34.1~GHz using $\beta=-2.24\pm 0.1$, from direct measurements (Mason et al.~1999).  The VSA calibration
is now expressed in terms of the WMAP  spectrum of Jupiter which is
tied to the CMB dipole temperature.  At 33  GHz, the most recent WMAP value is
T$_{J}$ = 146.6$\pm$0.75~K (Hill et al.~2008).\footnote{It is of interest to note that the previous WMAP value (Page et al.~2003) is identical to the new one (Hill et al.~2008) but with a larger error ($146.6\pm2.0$~K) due to beam uncertainties. This led the  CBI group (Readhead  et al. 2004) to use a weighted estimate between
the  Mason et al. (1999) and earlier WMAP values, obtaining T$_{J}$ = 147.3$\pm$1.8~K at 32~GHz.}  The
ultimate  absolute  precision of our adopted intensity scale is
therefore $\approx~0.5$~\%.  Of  course we  can, in principle, measure relative flux
densities/brightness temperatures more  accurately than this.

 Accurate determination of relative intensities of  sources is made
from the daily calibration observations and includes  data from both
the compact and the extended VSA configurations.  Longer term flux
density  or brightness temperature variability measurements are
possible from the  observations which extend from September 2000 to
October 2004.

\begin{figure} \centering 
\mbox{\resizebox{0.48\textwidth}{!}{\includegraphics[angle=0]{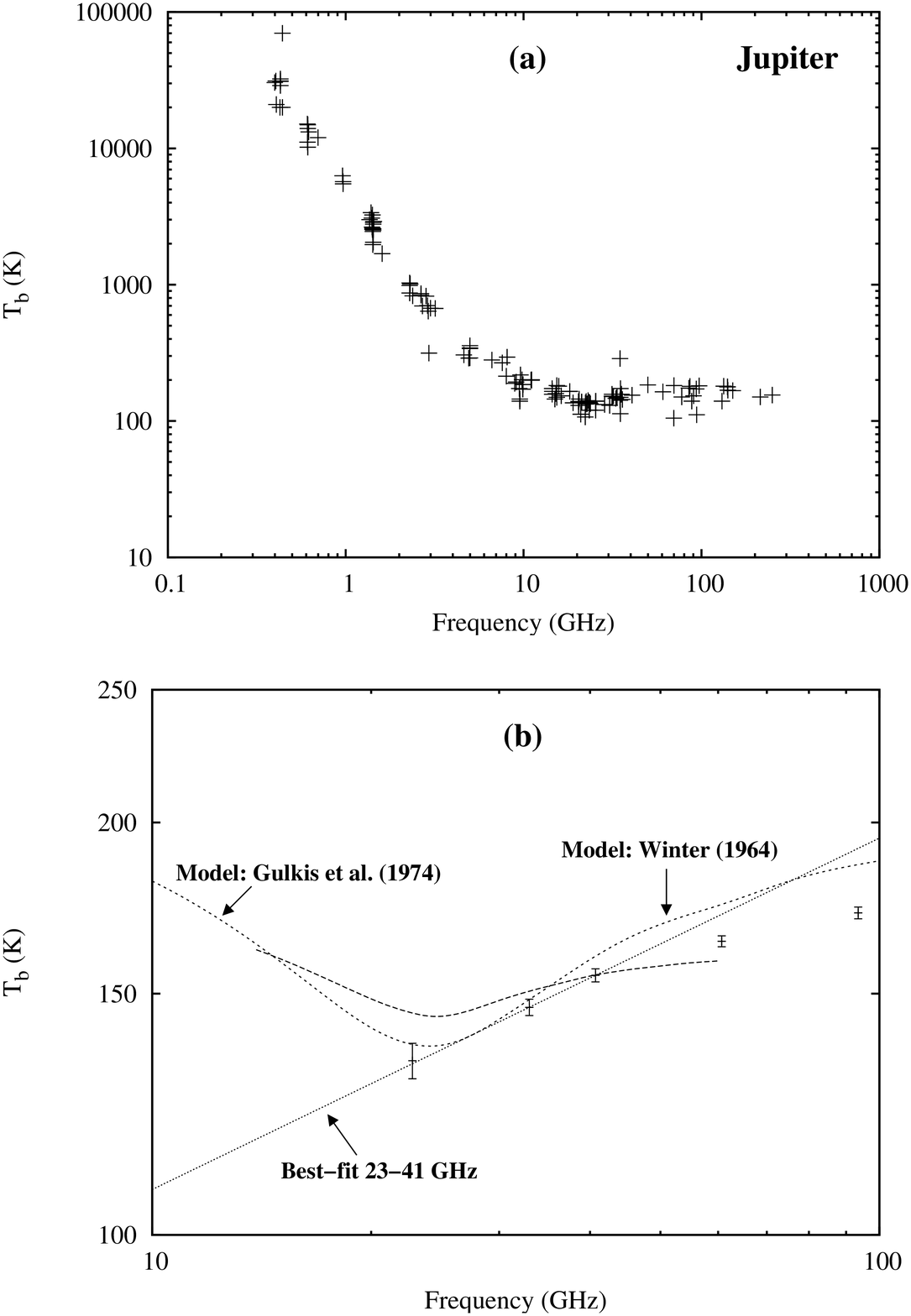}}} 
\caption{ The brightness temperature spectrum 
of Jupiter.(a) The spectrum covering 0.1 to 300~GHz from published data;  
cyclotron emission dominates the lower frequencies and thermal emission 
the higher frequencies. (b) The spectrum centred around 30~GHz showing the 
accurate observations from WMAP (Page et al. 2003).  Models by Gulkis et 
al. (1974) and Winter (1964) show the effect of ammonia absorption. For the model of Winter (1964), we used values of $T_{c}=168$~K and $P_{c}=2$~atm.}
\label{jupspec}
\end{figure}

The brightness  temperature spectrum of Jupiter is
shown between 0.1 and 200 GHz in  Fig. \ref{jupspec}.  The broad
depression at around $20$~GHz is believed to  be due to  molecular
absorption by ammonia.  Fig. \ref{jupspec} also shows this part  of
the spectrum  in more detail mainly delineated by the accurate WMAP
points (Page et al.  2003).  The best-fit temperature spectral index
at 33 GHz is $\beta =  +0.248\pm.004$.   Models for the Jupiter
spectrum by Gulkis et al. (1974)  and Winter (1964) give  an
indication of the expected shape in this  frequency range.

\subsection{The atmospheric contribution} The major effect of the 
atmosphere on the calibration process in the VSA is its variable
contribution to the total system noise which is subject to the
automatic  gain control (AGC).  The second smaller contribution is the
absorption of  source radiation on its passage through the atmosphere
to the antenna.   Both effects vary with the cosecant of the
elevation.  A detailed  discussion of the atmospheric contribution is
given separately by Davies  et al. (in prep.).

 The AGC system gives a 
reduction of the astronomical signal relative to the zenith value by a 
factor,
\begin{eqnarray}
f_{1} & = & [{\rm T}_{R} + {\rm T}_{atm(z)}]/[{\rm T}_{R} + {\rm T}_{atm(E)}] \\
      & = & 1/[1+({\rm T}_{atm(z)}({\rm cosec}E-1)/(T_{R} + {\rm T}_{atm(z)}))] \nonumber
\label{eq1}
\end{eqnarray}

where T$_{R}$ is the receiver noise plus cosmic background 
and 
T$_{atm(z)}$ and T$_{atm(E)}$ are the atmospheric emission at the zenith 
and 
at 
elevation $E$ respectively. For an average total set noise at the zenith of 
T$_{R}$ + T$_{atm(z)}$ = 35 K applicable to the VSA antennas and 
T$_{atm(z)}$ = 
7.5 
K, we obtain
\begin{equation}
f_{1} \approx 1-0.23 ({\rm cosec}E-1)
\label{eq2}
\end{equation}

 comprising 6.5 K of O$_{2}$ emission and 
1.0 
K 
water vapour emission, equivalent to 3mm precipitable water vapour. A 
similar factor relating the intensity of the astronomical signal arriving 
at the antenna from an elevation E to that which it would have been at the 
zenith is 

\begin{eqnarray}
f_{2} &  = & [e^{-\tau (E)}]/[e^{-\tau (z)}] \\
      &  = & 1-\tau (z)({\rm cosec}E-1). \nonumber
\label{eq3}
\end{eqnarray} 

For the relevant model atmosphere above Iza\~na 
the 
atmospheric temperature is 260 K, so with T$_{atm(z)}$ = 7.5 K,

\begin{equation}
f_{2} = 1-0.028({\rm cosec}E -1).
\label{eq4}
\end{equation} 

The final correction to the source observations for 
the atmosphere,  to first order  in $({\rm cosec}E-1)$, is 

\begin{equation}
f = f_{1}\times f_{2} = 1-0.26({\rm cosec}E-1). 
\label{eq5}
\end{equation}

Since the VSA is 
restricted to elevations above 50$^\circ$, the corrections are less than 
7\% 
and the approximations in the expressions for $f$, $f_{1}$ and $f_{2}$ are 
valid. In each pair of observations for a flux density comparison, 
each source is corrected for elevation as in equation (5) and the source 
ratio corrected to the zenith is obtained. 

\begin{figure} \centering 
\mbox{\resizebox{0.48\textwidth}{!}{\includegraphics[angle=0]{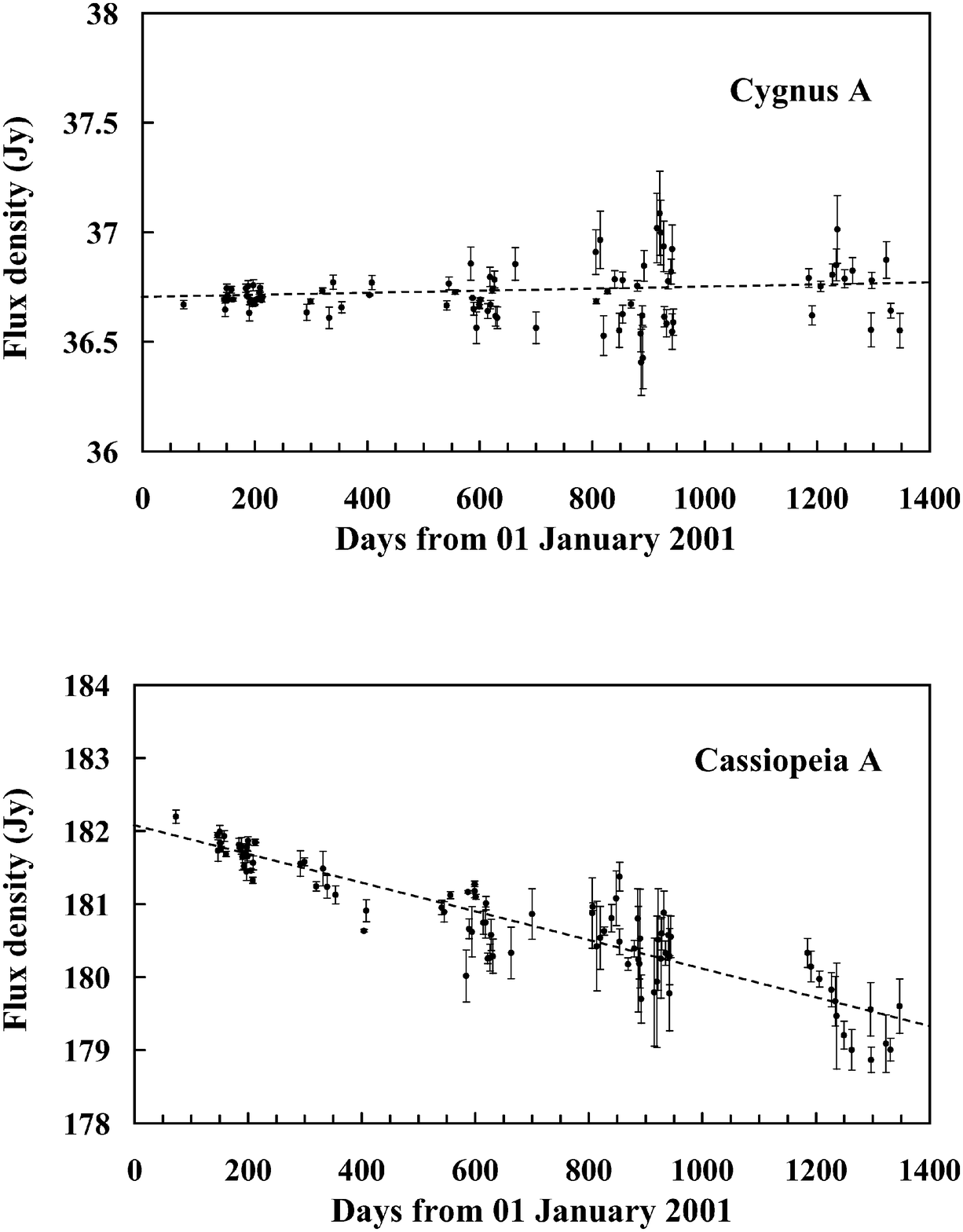}}} 
\caption{ VSA flux density measurements of Cyg~A and Cas~A calibrated by  
Jupiter over the period 13 March 2001 to 9 August 2004. The solid lines are least-square fits to the data.}
\label{cygcas}
\end{figure}

A convincing demonstration can be 
made of this calibration technique. The quality of the data corrected as described above can be assessed by 
intercomparing the 3 calibrators Jupiter, Cas~A and 
Cyg~A, all observed near transit and spread over the period 2002 to 2004. 
Fig.~\ref{cygcas} shows plots of the observed flux density of Cyg~A and Cas~A using 
Jupiter as a calibrator. A similar plot for Tau~A is given in Fig.~\ref{tausec} below. 
The daily scatter in the fluxes measured relative to Jupiter is seen to be 
$< 1$\% . The level of thermal noise in a typical daily measurement is $\sim 0.1$~Jy while the daily scatter is $\sim 1-2$~Jy, thus the variations are dominated by the atmosphere. In the case of Cyg~A and Cas~A, the time difference between the 
source and Jupiter  observation can be as large as 12hr. For Tau~A, where the 
time difference is smaller ($ < 6$hr) over this period, the scatter was 
$<0.5$\%.

Since the main contributor to the scatter is atmospheric water vapour 
emission, the 1\% scatter corresponds to 1\% of set noise within the AGC 
system, which is 0.35~K. At 33 GHz this temperature scatter corresponds to 1~mm pwv 
(Danese \& Partridge 1989). Some of the implications of this high level of 
stability in the atmosphere is included in the discussion of individual 
sources in sections 4.1 to 4.3.

\section{Derived source parameters} We now derive the flux density and/or 
brightness temperature of our calibration sources at 33.0 GHz relative to 
Jupiter.

 Since our results are most useful to the community working in the 
$\sim~10$~GHz wide band around 33 GHz, we have derived a spectral index 
for that range from flux densities in the literature (plus our new 
determinations). The brightness temperature spectrum of Jupiter has already been discussed in Section 3.3. We 
extrapolated all the VSA data to a common frequency of 33 GHz using the 
spectral indices derived from the literature.

 For several of the sources in our study, data are available for time 
spans between 1 and 4 years. These accurate data are used to investigate 
source variability at high precision.

\subsection{Cygnus A} Cyg~A is a radio source associated with a 
double galaxy at a redshift of $z = 0.056$ containing a central core and hot spots at the outer ends of diffuse radio 
lobes. The overall extent is 2.1 arcmin and is effectively 
unresolved in the compact and extended VSA arrays. No variability has ever 
been reported for Cyg~A at any wavelength.

\begin{figure} \centering 
\mbox{\resizebox{0.48\textwidth}{!}{\includegraphics[angle=0]{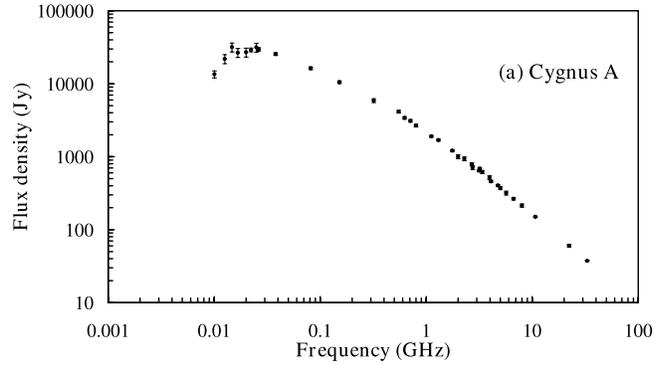}}} 
\caption{The flux density spectrum of Cyg~A from 10 MHz to 
100 GHz; 
the low frequency turnover is due to Galactic free-free absorption.} 
\label{cygspec}
\end{figure}

The radio spectrum of the integrated emission of Cyg~A taken from
literature is shown in Fig.~\ref{cygspec}.  There are several changes
of slope in the spectrum over the  frequency range 0.1 to 100 GHz; one
is at $\sim~0.5$~GHz and another is at  $\sim~3$~GHz. The slope of the
section 5-100~GHz, which includes 33~GHz,  is $-1.208\pm0.017$. We
recommend this as the spectral index at 33~GHz.  This value may be
compared with $-1.244\pm0.014$ given in Table~3 of Baars  et
al. (1977) at similar frequencies.

\begin{table} \centering 
\caption{Cassiopeia A flux density for epochs 1965 and 2000.} 
\label{casflux} \begin{tabular}{cccc} \hline\noalign{\smallskip} 
Frequency &S$_{\rm 1965}$ &S$_{\rm 2000}$ &Secular decrease\\
GHz  &Jy &Jy  &$\%$~per year\\
\hline\noalign{\smallskip}
1.405	 &2439$\pm$50	&1970$\pm$50     &0.67\\
1.415	 &2470$\pm$50	&1885$\pm$40     &0.67\\
1.44	 &2328$\pm$50	&1793$\pm$40     &0.66\\
1.44	 &2367$\pm$20	&1813$\pm$55     &0.66\\
3.15	 &1258$\pm$38	&1030$\pm$30     &0.62\\
3.2	 &1279$\pm$58	&1008$\pm$45     &0.61\\
4.08	 &1084$\pm$26	&860$\pm$20      &0.59\\
6.66	 &684$\pm$20	&548$\pm$16      &0.57\\
8.25	 &615$\pm$22	&497$\pm$18      &0.55\\
10.7	 &468$\pm$0	&388$\pm$0       &0.54\\
13.49	 &394$\pm$13	&323$\pm$11      &0.53\\
14.5	 &367$\pm$10	&310$\pm$9       &0.52\\
15.5	 &376$\pm$18	&309$\pm$15      &0.51\\
16	 &354$\pm$11	&292$\pm$9       &0.51\\
22.28	 &285$\pm$10	&236$\pm$9       &0.49\\
32	 &224$\pm$6	&192$\pm$5       &0.47\\
33	 &211$\pm$5	&183$\pm$5       &0.47\\
86	 &115$\pm$4	&100$\pm$4       &0.41\\
87	 &109.4$\pm$0	&95.4$\pm$0      &0.41\\
140	 &78.3$\pm$7	&69.1$\pm$6.2    &0.38\\ 
250	 &51.8$\pm$5.6	&47.2$\pm$5.3    &0.36\\
\noalign{\smallskip}\hline
\end{tabular}
\end{table}

Observations of the calibration sources Cyg~A, Cas~A, and Jupiter were 
analysed for the period March 14 2001 to 4 May 2004. The flux density of 
Cyg~A calibrated by Jupiter (assumed to have a brightness temperature of 
146.6 K at 33 GHz) over this period is shown in Fig \ref{cygcas}. In this 
period the RA of Jupiter ranged from RA=08h to 12h so observations were 
typically separated by 8-12 hr. The 1\% scatter in the derived flux 
density of Cyg~A was due to the atmosphere (mainly water vapour) over this 
time range. The mean flux density at 33 GHz was found to be 36.4$\pm$0.2 
Jy. 

These data can also be used to probe the long-term stability of the radio 
emission from Cyg~A and Jupiter at 33~GHz. A fit to the data shows that the 
flux density ratio of Cyg~A and Jupiter increases by 0.043$\pm$0.039\% per 
year. Accordingly we conclude that the relative radio emission changed by 
$\leq$ 0.1\% per year over the period March 2001 to May 2004.

\subsection{Cassiopeia A} 

Cas~A is a 330-year old shell-type SNR of $5$~arcmin diameter. Baars et al. (1977) give its spectral index between 0.3 
and 31 GHz as $\alpha = -0.792\pm0.007$ in 1965 and $\alpha = -0.770\pm0.007$ in 
1980. Radio imaging shows multiple hot spots with spectral indices 
between $-0.65$ and $-0.90$ (see e.g. Anderson \& Rudnick 1996) embedded in extended 
structure. Steeper spectra are associated with features thought to be 
bowshocks and with features outside the main radio ring; flatter 
spectra are found in the ring and in bright features within it which may  
account for its slow but measurable decrease in flux density with time at 
a rate which is thought to vary with frequency. These are important 
considerations when using Cas~A as a calibrator.

\subsubsection{The Cas~A spectrum} 

\begin{figure}
\centering
\mbox{\resizebox{0.48\textwidth}{!}{\includegraphics[angle=0]{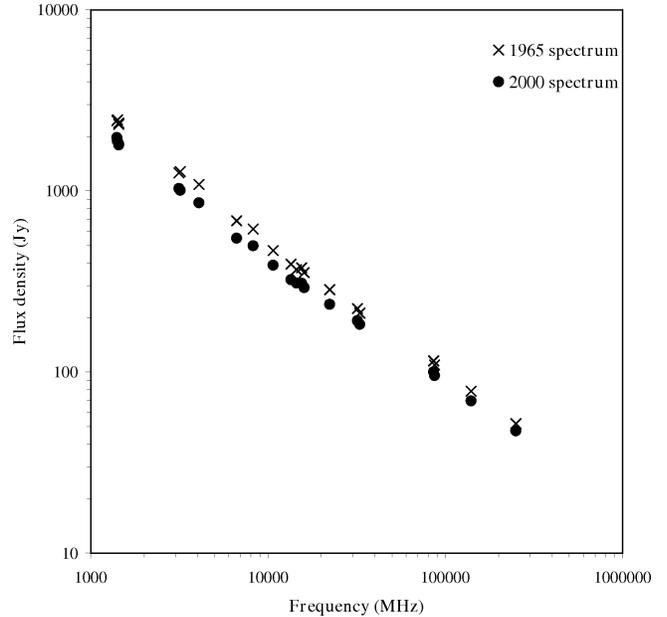}}}
\caption{ The spectrum of Cas~A for the epochs 1965 (crosses) and 2000 (filled 
circles) calculated from published data with the secular decreases applied 
appropriate to each frequency (see Section 4.2.3).
} 
\label{casspec}
\end{figure}

Fig.~\ref{casspec} shows the spectrum of  Cas~A taken from published
data over  the frequency range 1.0 to 100 GHz with the flux densities
corrected to  the years 1965 and 2000 using the recipe given in
Section~4.2.3 and listed in  Table~\ref{casflux}. The spectrum  over
this large frequency range shows a flattening of the spectral index
above 15 GHz  not detected in earlier data; this flattening is seen in
both the 1965 and 2000 data.  In 1965 $\alpha = -0.795\pm0.12$ at $\nu
< 15$ GHz and $\alpha = -0.725\pm0.013$ at  $\nu >15$ GHz. For the
2000 data $\alpha = -0.767\pm0.018$ at $\nu < 15$ GHz and  $\alpha =
-0.663\pm0.012$ at $\nu > 15$ GHz. An appropriate spectral index at 33
GHz  at the present epoch would appear to be $\alpha = -0.69\pm0.02$.

\subsubsection{VSA measurements of Cas~A} 
\begin{figure}
\centering
\mbox{\resizebox{0.48\textwidth}{!}{\includegraphics[angle=0]{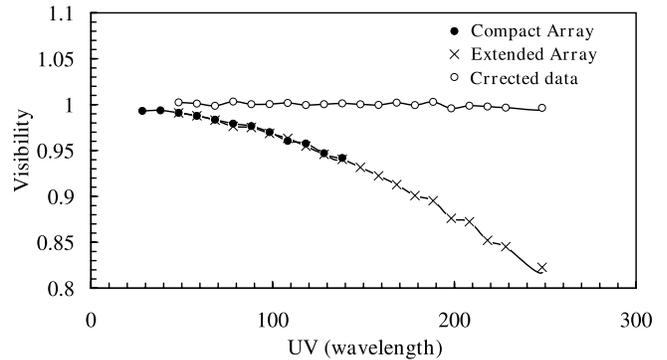}}}
\caption{ The visibility amplitude of Cas~A as a function of VSA baseline 
length 
measured in wavelengths. The crosses show the visibilities measured with 
the extended array; the filled circles show the compact array. The open 
circles  are the visibilities corrected to a 5 GHz observed map of Cas~A; 
this 
correction brings the visibility  to within 1\% of unity.} 
\label{cascorr} 
\end{figure}

With the resolution of 17 and 11 
arcmin for the compact and extended forms of the VSA, Cas~A will show 
significant resolution effects in a project which aims for an accuracy of 
better than 1\%. Accordingly, a correction was applied in the 
analysis procedure which scaled the visibilities on each baseline. The 5~GHz map from the VLA was used for this purpose. In view of the differing 
spectral indices within Cas~A, it was necessary to confirm that the 5~GHz 
model was applicable to our 33 GHz observations. Fig.~\ref{cascorr} 
which plots the visibility amplitude of Cas~A versus the baseline length 
shows the 
results of this investigation. It can be seen that the normalised 
visibility after correction with the model is unity within 1\% across the 
baseline range 20 to 250 $\lambda$. The compact and extended VSA data sets 
overlap at a precision of better than 0.5\% in the baseline range 50-130 
$\lambda$. The correction model is thus adequate for our purpose. Fig.~\ref{cygcas}(b) shows the flux density of Cas~A measured over the period 13 
March 2001 to 9 August 2004. Each point is the average of the flux 
densities measured relative to Cyg~A and Jupiter. The secular decrease in 
the flux density of Cas~A is clearly seen. No seasonal effects are evident 
in the data as the 3 sources move from day to night. Moreover the data for 
the compact array (March 2000 to May 2001) are entirely consistent with 
those from the extended array; 
no step is seen in the plot around May 2001. A linear fit to the data of 
Fig.~\ref{cygcas}(b) gives a flux density for Cas~A on the adopted Jupiter 
scale of 182.02$\pm$0.07~Jy on January 2001. The linear rate of decrease 
over the period March 2001 to August 2004 is given by

\begin{equation}
\frac{{\rm d}S}{S{\rm d}t}  = -0.394 \pm 0.019 \%~{\rm yr}^{-1}.
\label{eq6}
\end{equation}

\subsubsection{Secular decrease in flux density of Cas~A} 

The current 
estimate of the secular decrease in the flux density of Cas~A at 33 GHz 
obtained over a 3.5 year period in the present decade is the most accurate 
of any epoch. Scott et al. (1969) at 81.5~MHz covering the two decades $1949-1969$ found a value of $1.29\pm0.08$\%~yr$^{-1}$. 
Using more recent data at 81.5 MHz, Hook et al. (1992) found a lower value rate of decrease between 1949 and 1989.7 of $0.92\pm0.16$\%~yr$^{-1}$.

 We now re-examine the data available 
relating to the variability of Cas~A as a function of both epoch and 
frequency. The data between 1949 and 1976 in the frequency range 81.5 MHz 
to 9.4 GHz indicated a secular decrease of flux density estimated by Baars 
et al. (1977) to be frequency dependent and given by the relation 

\begin{equation}
\frac{{\rm d}S}{S{\rm d}t} = 0.97 (\pm 0.04) -0.30 (\pm 0.04) log (\nu/GHz)\%~{\rm yr}^{-1}. 
\label{eq7}
\end{equation}

The most 
accurate data currently available covers the frequency range 38 MHz to 33 
GHz and the period 1949 to 2004 is given in Table \ref{seccas}.

\begin{table}
\centering 
\caption{Secular decrease of Cas~A as a function of frequency.}
\label{seccas} 
\begin{tabular}{ccccccccccc}
\hline\noalign{\smallskip}
Frequency &Epoch & Decrease \\
    GHz &  & \%~per year \\
\hline\noalign{\smallskip}
0.038	&1955-87 	&0.8$\pm$0.08\\	   
0.0815	&1949-69       &1.29$\pm$0.08\\	
0.0815    &1949-1989      &0.92$\pm$0.16\\  
0.0815	&1965-1989	&0.63$\pm$0.06\\	   
0.1025	&1977-1992 	&0.80$\pm$0.12\\	   
0.927	&1977-1996	&0.73$\pm$0.05\\	   
0.950	&1964-1972	&0.85$\pm$0.05\\	   
1.405	&1965-1999	&0.62$\pm$0.12\\	   
1.420	&1957-1976	&0.89$\pm$0.02\\	   
1.420	&1957-1971	&0.89$\pm$0.12\\	   
3.000	&1961-1972	&0.92$\pm$0.15\\	   
3.060	&1961-1971	&1.04$\pm$0.21\\	   
7.800	&1963-1974	&0.7$\pm$0.1\\	   
9.400	&1961-1971	&0.63$\pm$0.12\\   
15.500	&1965-1995	&0.6$\pm$0.06\\	   
33.000	&2001-2004	&0.39$\pm$0.02\\
\noalign{\smallskip}\hline
\end{tabular}
\end{table}

\begin{figure} \centering 
\mbox{\resizebox{0.48\textwidth}{!}{\includegraphics[angle=0]{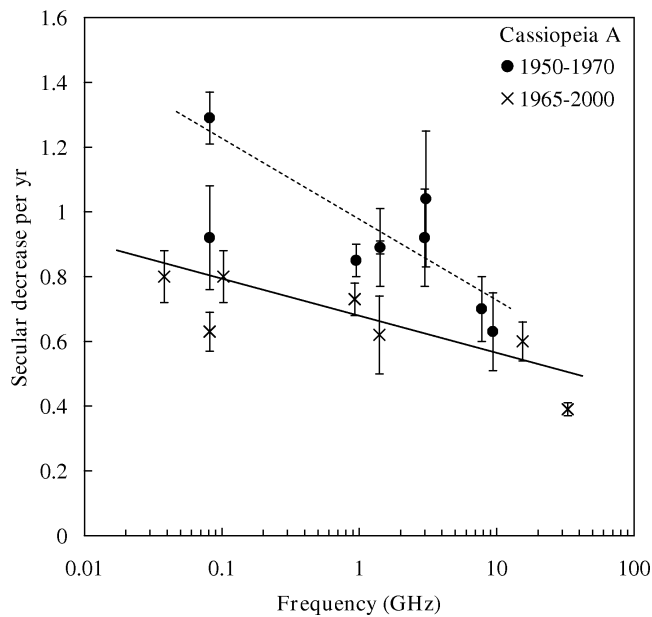}}} 
\caption{ The secular decrease in the flux density of Cas~A as a function 
of frequency for two epochs, 
1950-1970 (filled circles) and 1965-2000 (crosses). The lines are least-square fits to the data at the two epochs.} 
\label{casepoch}
\end{figure}

Fig.~\ref{casepoch} shows the data from Table \ref{seccas} divided into 
two epochs. 
The filled circles 
show the secular decrease from $\sim~1950$~to 1970 while the crosses 
show the decrease from $\sim~1970$~to 1990/2000. It is clearly evident 
that the earlier data shows higher ($\sim$ 50\%) secular decrease than the 
later data. The data from 1970 to 1990/2000 indicate a secular decrease of

\begin{equation}
\frac{{\rm d}S}{S{\rm d}t} = 0.68 (\pm 0.04 ) -0.15 (\pm 0.04) {\rm log} 
(\nu/{\rm GHz})\%~{\rm yr}^{-1}.
\label{eq8}
\end{equation} 

The material available is not sufficient to show whether the 
fall in the secular rate is linear with time or is or not (as 
proposed for the 81.5 MHz data by Hook et al. (1992)). Apart from the 
current (2001-2004) 33 GHz result the values given by equation 6 may be an 
upper limit to the secular variation of Cas~A at the present epoch. The secular decrease of Cas~A will be discussed further in a related paper (Lancaster et al., in prep.).

 A frequency-dependent secular variation has implications at the 
present epoch for the spectrum of Cas~A. We can apply equation (7) 
to the higher weight data from Table 2 of 
Baars et al. (1977) and more recent data particularly at higher 
frequencies (Reichart \& Stephens 2000, Mezger et al. 1986, Mason et al. 
1999, Liszt \& Lucas 1999, Wright et al. 1999, present paper) to derive 
spectra for 1965 (the Baars et al., epoch) and for 2000 at frequencies 
from 1 GHz to 250 GHz). The decrease of the secular term with frequency as 
shown by equation (7) gives a flattening of the spectral index with time.  
A synchrotron 
spectral index which flattens with time implies that particle acceleration 
is continuing in Cas~A, presumably at the shock interfaces in the object 
(see e.g. Reynolds \& Ellison 1992).

\subsection{Taurus A} 

The Crab Nebula is the filled-centre remnant of 
the 
supernova of 1054 AD. The central neutron star remains active and replenishes 
the supply of relativistic electrons in the nebula. There is some evidence 
for a small secular decrease of the flux density at both radio and optical 
wavelengths. The present observations lead to a flux density of Tau~A at 
33 GHz. Since the source is strongly linearly polarized we 
must determine the polarization in order to use Tau~A as a calibrator at 
1\% accuracy. 
\subsubsection{The spectrum of Tau~A} 
\begin{figure}
\centering
\mbox{\resizebox{0.48\textwidth}{!}{\includegraphics[angle=0]{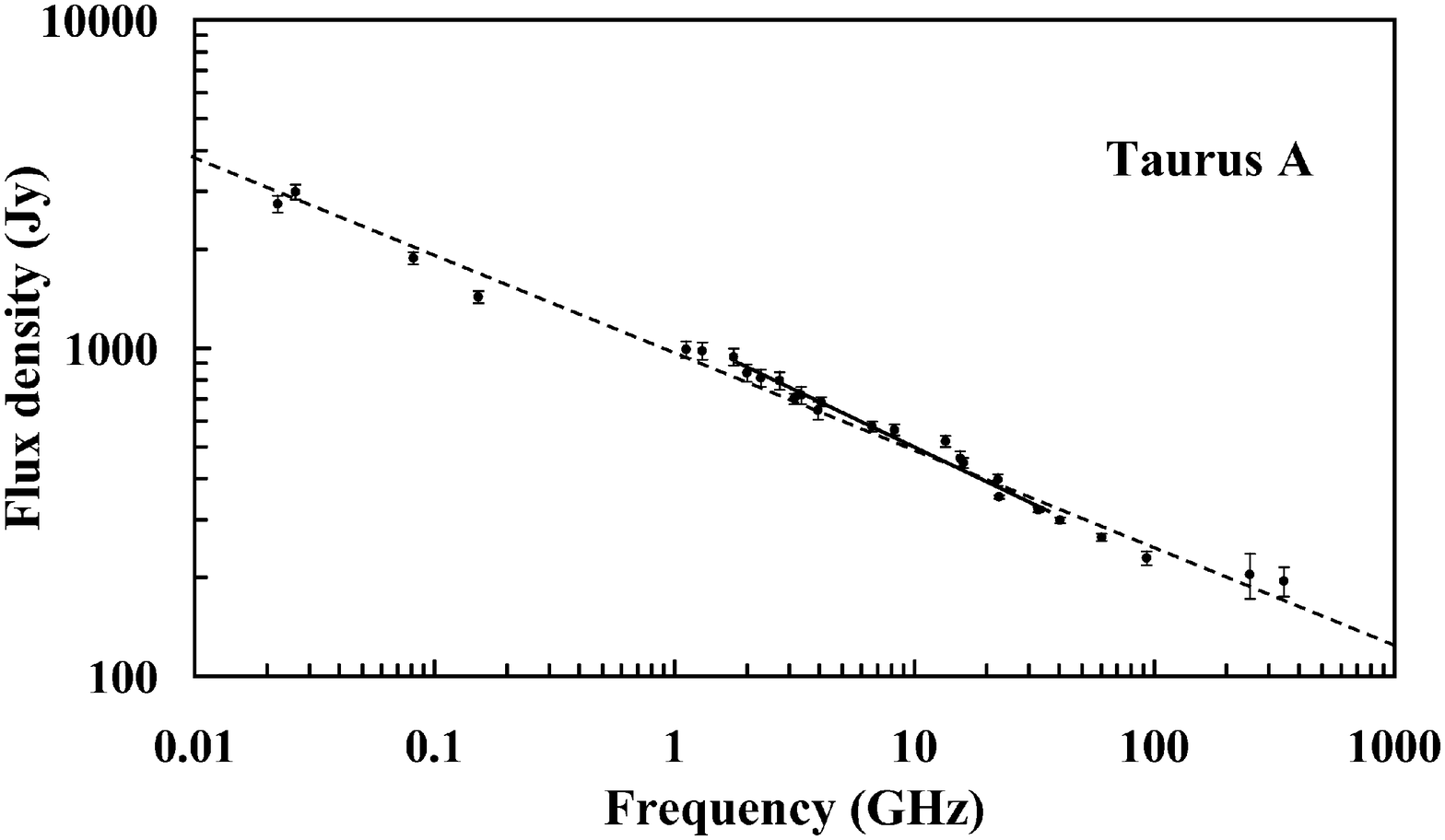}}}
\caption{ The log-log flux density spectrum of Tau~A from 10 MHz to 1000 GHz. A 
linear slope of $\alpha = -0.278\pm0.007$ fits the data from 1 to 350 GHz (dashed 
line).The slope between 1.8 and 33 GHz (full line) is $\alpha = -0.351\pm0.016$.} 
\label{tauspec}
\end{figure}

Baars et al. (1977) 
found a constant spectral index over the frequency range 1-35 
GHz with value of $\alpha = -0.299\pm0.009$. More recent data, including 
higher frequencies up to 350 GHz are shown in Fig.~\ref{tauspec}. A 
constant 
spectral index of $-0.278\pm0.007$ fits the data between 1 and 350 GHz. 
The expected turn-down of the spectrum is beyond this frequency range and 
is believed to be at 10 THz (Marsden et al. 1984). There is, however, 
some evidence 
for a small curvature of the spectrum.  We find $\alpha$ = 
$-0.258\pm0.028$ between 22 and 1800 MHz, $\alpha = -0.351\pm0.016$ 
between 1.8 and 33 GHz and $\alpha = -0.35\pm0.03$ between 8 and 347 
GHz. Green (2002) gives the spectral index between 1.5 and 347 GHz of 
individual structures in the nebula to lie in the range $-0.31$ and $-0.34$. 
Similarly Mezger et al. (1986) give a spectral index between images at 
10.7 and 250 GHz of $\alpha = -0.30\pm0.05$. We adopt $\alpha = 
-0.32\pm0.01$ as an appropriate spectral index at $\approx 30$~GHz, which is consistent with the recent analysis by Macias-Perez et al. (2008).

 \subsubsection{The VSA observations of Tau~A} 

Care is required when using  Tau~A as a calibrator in two
regards. Firstly, it is an extended source  with dimensions of 4 x 6
arcmin$^{2}$ at position angle (pa)  $\sim~140^\circ$ and consequently
requires a resolution correction when  observed with the VSA in both
the extended and compact configurations.  Secondly, it is some $10\%$
linearly polarized at frequencies of 30 GHz  and above and this must
be taken into account when observing with  linear polarization.

 The observed optical polarization of Tau~A is $9.2\%$ at
pa=159.6$^\circ$  (Oort and Walraven 1956).  The integrated radio polarisation
increases from 2\% at 1.4 GHz to 7\% at  5 GHz (Gardner \& Davies
1966). The mean Faraday rotation measure (RM) in  Tau~A is $-25$ rad
m$^{-2}$ with values up to 300 rad m$^{-2}$ in some  places. Any RM in
this range leads to negligible Faraday rotation (FR) 30~GHz (a
wavelength of 1.0 cm) corresponding to a FR$\leq$  1$^\circ$
(FR$\propto 1/\lambda^2$).

We now consider the VSA data for Tau~A.
\begin{enumerate} \item{The linear polarization of Tau~A}. This can be estimated from long HA tracks of the
source,  remembering that the VSA is linearly polarized at
pa=0$^\circ$. The  maximum HA  track for Tau~A runs from HA=$-3$h 00m
to $+3$h 00m. The observed data from  the extended array on 26
December 2001 calibrated against Cas~A (and  Jupiter) is shown in
Fig.~\ref{tauflux} by crosses. This data set is  first  corrected
(circles) for resolution effects of the VSA as for Cas~A  (section
4.2.2) using a 5 GHz map of Tau; this model was confirmed by  showing
that the corrected visibilities were constant with baseline for  both
the compact and extended VSA configurations in the manner  illustrated
for Cas~A in Fig.~5. The elevation correction  for absorption in the
atmosphere and the AGC effect on the receivers (0.26  cosec$E$ - see
Section 3.4) were then applied giving the true variation of  flux
density with HA shown as filled circles in Fig.~\ref{tauflux}. The
linearly  polarized flux density P is then estimated by fitting the
corrected data  to P cosec (2($p-\theta$)), where $p$ is the
polarization pa measured  east from  north and $\theta$ is the
parallactic angle at the HA of observation. Data  from 4 nights (26
December 2001, 09 February 2002, 12 October 2002 and  25 January 2003)
of good weather and  analysed as above were averaged to determine the
polarization of Tau~A.     The results of this fitting gave P =
24.6$\pm$1.6 Jy corresponding to  7.8$\pm$0.6\% linear polarisation at
pa=148$^\circ$$\pm$3$^\circ$.  The errors quoted include the
uncertainty in the atmospheric correction.

\begin{figure} \centering 
\mbox{\resizebox{0.48\textwidth}{!}{\includegraphics[angle=0]{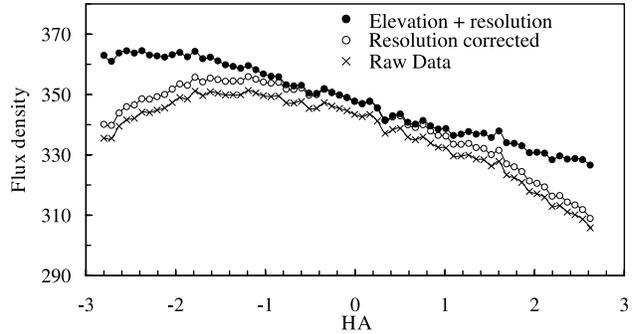}}} 
\caption{ The flux density of Tau~A plotted against HA for the observation on 26 
December 2001 with the extended array. Crosses show  the observed data calibrated 
against Jupiter; open circles show the data then corrected for resolution 
affects; filled circles show the data further corrected for elevation 
(atmospheric absorption and the AGC effect). The resultant sloping HA plot is due 
to the linear polarization of Tau~A.} 
\label{tauflux} 
\end{figure}

\item{The flux density of Tau~A at 33 GHz}. The total flux density of Tau~A is
determined by the fitting procedure given in the previous section.  The 
average of the 4 HA scans gives S(Tau~A) = 322 $\pm$4 Jy. The flux density 
of 
Tau~A at HA = 0hr (pa = 0$^\circ$) is 332 $\pm$4 Jy.
A more extensive data set using short calibration tracks running from 3 
January 2002 to 11 September 2004 with the extended array gave a more 
accurate 
estimate. The individual data points were corrected for resolution and for 
the 
atmosphere and are plotted in Fig.~\ref{tausec}.  The use of Jupiter as the reference for Tau~A led to the very low 
scatter (rms = 0.4\%) in the data since they were both night-time objects 
and separated in time by less than 5 hrs.  This data set gives a flux 
density for Tau~A at pa = 0$^\circ$ of
332.8$\pm$ 0.7 Jy on 1 January 2001.  The secular decrease of Tau~A 
indicated by Fig.~\ref{tausec} is discussed in the next section. 
\end{enumerate}

\subsubsection{Secular decay of Tau~A} The 950-year old nebula is clearly 
in the decay phase although centrally reenergised from the spindown of
its central pulsar. Theoretical
models predict a secular decrease of 0.16 to 0.4\% per year at radio
wavelengths (Reynolds and Chevalier 1984). An extended series of
observations at 8.0 GHz over the period 1969-1985 by Aller \& Reynolds
(1985) gave a decrease of 0.167$\pm$0.015\% per year, a value consistent 
with the theoretical models (an interesting determination of the optical
decay of the continuum at 5000 Angstrom using data from 1960 to 2002.2 
has been  
made by Smith (2003) who found a decay of 0.5$\pm$0.2\% per year).

 The current 33 GHz observations of Tau~A presented in Fig.~\ref{tausec}
 and
described in the previous section give a secular decrease over the period
January 2002 to September 2004 of

\begin{equation}
\frac{{\rm d}S}{S{\rm d}t} = -0.22 \pm0.07 \%~yr^{-1}.     
\label{eq9}
\end{equation}

This is very similar to the value obtained by Aller \& Reynolds (1985) at
8.0
GHz for the longer but earlier period 1968 to 1985 and is also consistent
with the models of Reynolds \& Chevalier (1984). A secular decrease of
$\sim~0.2$~\% per year would seem to be appropriate for the present epoch
at 33 GHz.  

\subsection{NGC7027} NGC7027 is young planetary nebula which is 
optically 
thick in Lyman continuum and is therefore an ionization-bounded ionized 
hydrogen region. The radio continuum and optical nebula have dimensions of 
12 x 8 arcsec$^{2}$ (Basart \& Daub 1987). It is surrounded by a ring of 
neutral molecular gas and dust some 40 arcsec in diameter possibly 
extending to 70 arcsec (see e.g. Hoare et al. 1992; Bieging et al. 1991). The 
expansion age of NGC7027 is 1200~yr with an expansion rate of 
4.2$\pm$0.6 mas per year (Masson 1989).

 Three factors are of particular interest in the 
present study. The first is the activity within the nebula; its expansion 
velocity is 17.5$\pm$1.5 kms$^{-1}$ with local gas velocities up to 55 
kms$^{-1}$ 
(Lopez et al. 1998, Cox et al. 2002). These motions may lead to secular 
variations of radio flux density. The second effect is the clumpiness of 
the ionized gas in NGC7027 which results in optically thick knots up to 
at 
least 5 GHz (Bains et al. 2003). This determines the frequency range for 
which an optically thin radio thermal spectrum (at T$_{e}\approx$14000~K) may 
be adopted. Thirdly, the dust within and surrounding NGC7027 may produce 
anomalous radio emission as in the case of the Helix Nebula (Casassus 
et al. 2004) and LDN 1622 (Casassus et al. 
2006). Such emission, which peaks in the range 10-30 GHz (Draine \& 
Lazarian 1998), needs to be considered when discussing the spectrum 
of 
NGC7027.

\begin{figure} \centering
\mbox{\resizebox{0.48\textwidth}{!}{\includegraphics[angle=0]{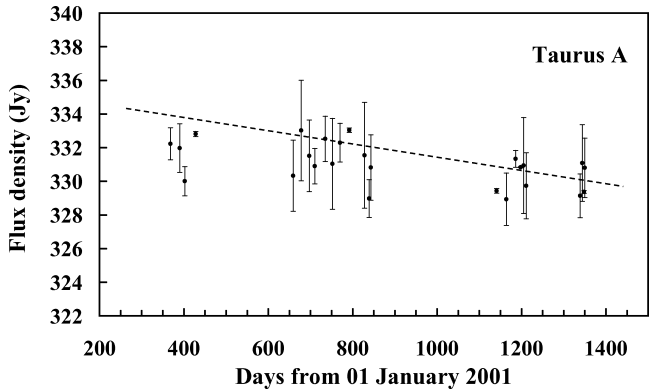}}}
\caption{ The flux density of Tau~A measured relative to Jupiter for the period 3 
January 2002 to 11 September 2004. The dashed line is a least-square fit to the data. The observed secular change is 
$-0.22\pm0.07\%~yr^{-1}$.} 
\label{tausec}
\end{figure}

\subsubsection{Spectrum of NGC7027} 
\begin{figure}
\centering
\mbox{\resizebox{0.48\textwidth}{!}{\includegraphics[angle=0.0]{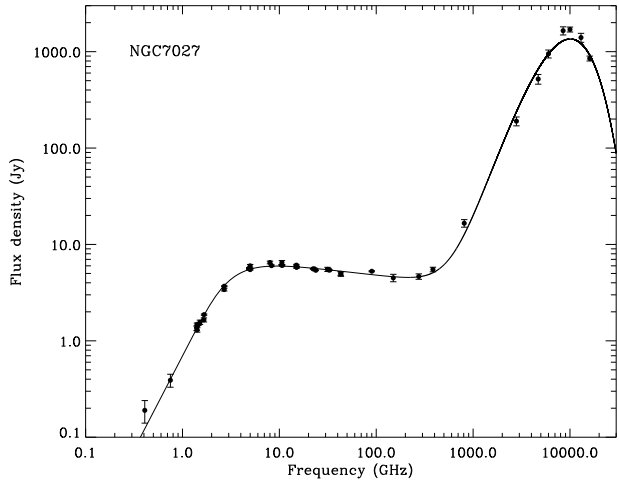}}}
\caption{ The flux density spectrum of NGC7027 covering the range 408~MHz to 16000~GHz using data from the literature. The solid line is a fit consisting of a optically thick and thin free-free spectrum in addition to a single component modified black-body dust model. At frequencies of $\sim 30$~GHz, the spectrum is dominated by optically thin free-free emission.} 
\label{NGC7027spec} 
\end{figure}

The radio spectrum of the integrated 
emission from NGC7027 is shown in Fig.~\ref{NGC7027spec} from the radio to 
the far 
infrared 
(FIR). It consists of several components. At radio frequencies the 
spectrum is due to free-free emission, optically thick at the lower 
frequencies and optically thin at the higher radio frequencies. The 
transition to optically thin occurs at $\sim$5-10 GHz in agreement 
with the observation by Bryce et al. (1997) that the bright NW knot has an 
optical depth of $\sim$2.0 at 5~GHz. A fit to the data between 10 and 
80 
GHz gives an optically thin spectral index of $-0.119 \pm 0.025$. The 
expected 
local value of the thermal spectral index for the observed electron 
temperature (14,000 K) in this frequency range lies between $-0.111$ (at 10~GHz) and $-0.133$ (43~GHz) (Dickinson et al. 2003).

 The FIR emission consists 
of two components. One is warm (100 K) dust inside 
the $12 \times 8$~arcsec$^{2}$ ionized 
hydrogen region and the other is cool (30 K) dust in a neutral gas ring 40 
arcsec in diameter surrounding the nebula (Hoare et al. 1992, Bieging et 
al. 1991). Both these dust components are covered by the observing 
beamwidth of the VSA
and are included in the integrated emission described here.

 The question 
of whether there is significant anomalous radio emission from either or 
both of the dust components can be addressed. There is no obvious emission 
with a peaked spectrum of roughly an octave width centred at $\approx 20$~GHz 
in the spectrum of Fig.~\ref{NGC7027spec}. For a normal cool dust cloud the expected 
emission based on the 
100$\micron$ (3~THz) 
flux density would be $\approx 0.3$~Jy (Davies et al. 2006, Watson et al. 
2005, Casassus et al. 2006). The emission from a normal HII 
region (Dickinson et al. 2006 \& 2007) would be $\approx 0.1$~Jy.  The latter 
is more 
consistent with Fig.~\ref{NGC7027spec}.

\begin{figure}
\centering
\mbox{\resizebox{0.48\textwidth}{!}{\includegraphics[angle=0]{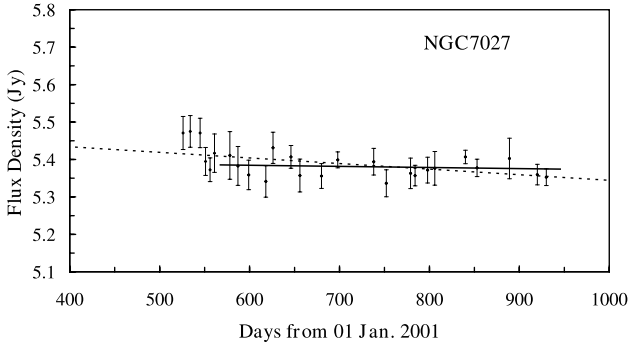}}}
\caption{ The flux density of NGC7027  calibrated by Jupiter and Cyg~A for 
the 
period June 2002 to July 2003, as measured by the VSA. There is weak evidence for a higher flux density 
in the first 6 data points. The annual change for the full data set (dashed 
line) is $-0.8\pm0.3\%~yr^{-1}$; the change for the data excluding the first 6 
points (full line) is $-0.31\pm0.37\%~yr^{-1}$.} 
\label{ngcshort} 
\end{figure}

 \subsubsection{The variability of NGC7027} 
The flux densities measured over the period June 2002 to July 2003 are 
shown in Fig.~\ref{ngcshort}. There is weak evidence for a higher flux density in 
the 
6 observations before 10 July 2002. After this time the flux density 
remains constant for the year; a linear fit to the latter data gives an 
insignificant annual change of $-0.31\pm0.37$\% yr$^{-1}$. A fit to the 
full run of the data gives a decrease of 0.8$\pm$0.3\% per year. The 
average flux density of NGC7027 in the period June 2002 to July 2003 is 
5.39$\pm$0.04 Jy. 

\begin{figure}
\centering  
\mbox{\resizebox{0.48\textwidth}{!}{\includegraphics[angle=0]{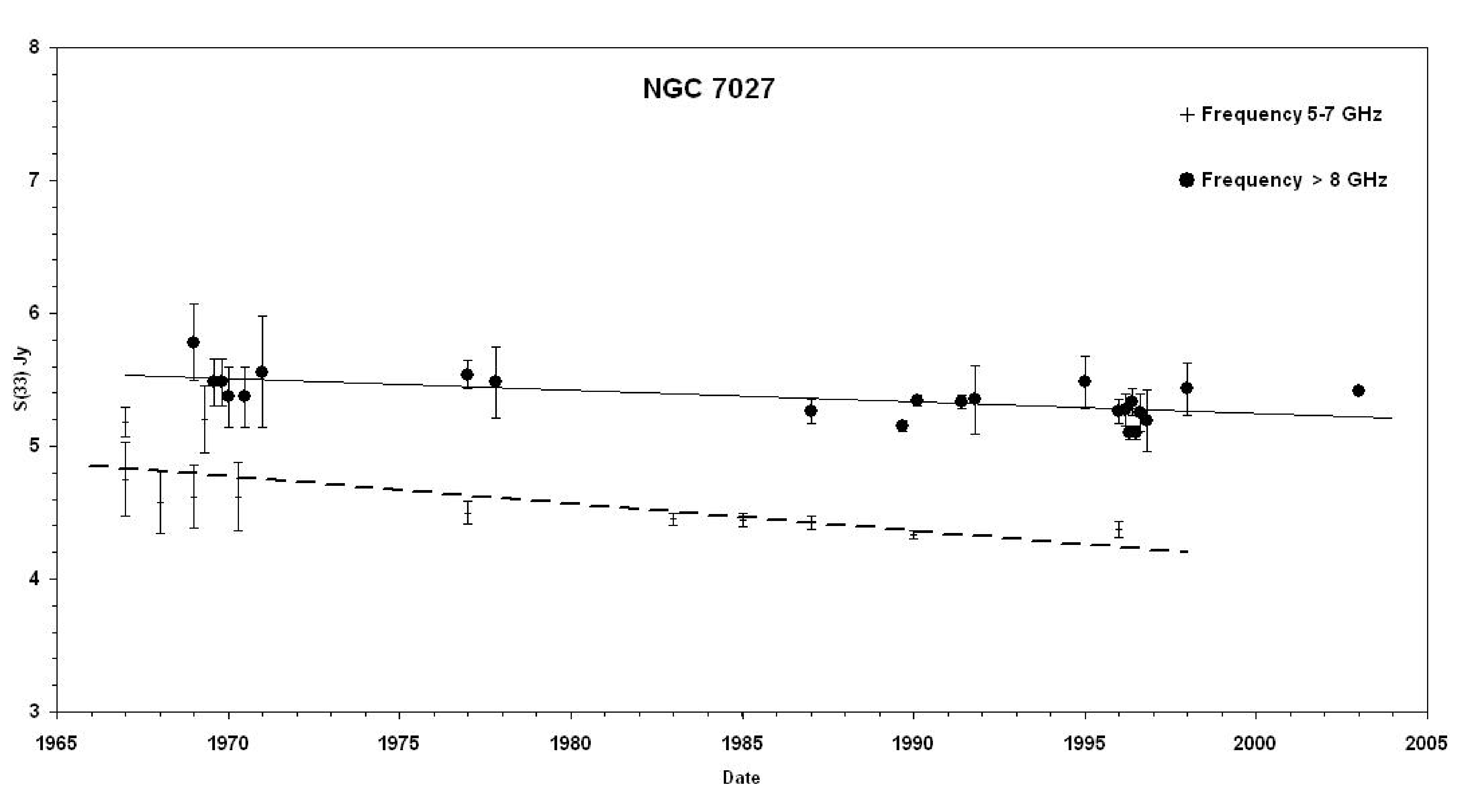}}}
\caption{ The flux density of NGC7027 from published data for two 
frequency 
ranges over the period 1967 to 2004 with least-square fits. \newline (a) Full line and filled circles; 
optically thin frequencies, 8.0 to 43~GHz, and \newline (b) dashed line and 
crosses; partially optically thick frequencies, 4.8 to 6.6~GHz.} 
\label{ngclong} 
\end{figure}

The long-term secular variation of NGC7027 at radio 
frequencies is often discussed (see e.g. Ott et al. 1994; Masson 1989).
A substantial data set covering the frequency range 4.8 to 43~GHz is 
available in the literature extending back to 1967.  Fig.~\ref{ngclong}
 shows the 
flux density of NGC7027 from 1967 to 2003 in two frequency ranges: $4.8 - 
6.6$~GHz where free-free self-absorption becomes significant and $8.0 - 43$~GHz where the spectrum is optically thin as can be seen in 
Fig.~\ref{NGC7027spec}.  
All flux densities are corrected to a frequency of 33 GHz, assuming an 
optically thin spectral index $\alpha = -0.12$ as expected theoretically for 
this frequency range (Dickinson et al. 2003).  The 4.8 - 6.6 GHz flux 
densities (median frequency 5.2 GHz) are 82 $\pm$ 3\% of the  optically 
thin 
frequencies, indicating $\tau$=0.18 at 5 GHz. Giving each point equal 
weight, the secular change in the flux density at optically thin 
frequencies (8.0 to 43 GHz) is

\begin{equation}
\frac{{\rm d}S}{S{\rm d}t} = -0.159 \pm0.043 \%~yr^{-1}.
\label{eq10}
\end{equation}

At the slightly optically thick frequencies (4.8 - 6.6 GHz)

\begin{equation}
\frac{{\rm d}S}{S{\rm d}t} =-0.45\pm 0.13\%~yr^{-1} 
\label{eq11}
\end{equation}

when taking all the data.  
However, discarding the earlier data with larger errors, the secular 
change 
from 1975 to 1996 is

\begin{equation}
\frac{{\rm d}S}{S{\rm d}t} = -0.183\pm0.052 \%~yr^{-1} 
\label{eq12}
\end{equation}

in close agreement with the optically 
thin data.  We therefore have consistent results from two independent 
data sets. On combining the above data, we conclude that for NGC7027

\begin{equation}
\frac{{\rm d}S}{S{\rm d}t} = -0.17\pm0.03 \%~yr^{-1} 
\label{eq13}
\end{equation}

over the period $\sim$1970 to the present. 

After completion of this work Zijlstra, van Hoof \& Perley (2008) 
published observations made with the VLA at frequencies in the range 1.275 
to 43.34 GHz covering the period 1983-2006. They find that over this 
shorter period the flux density increases at lower frequencies where the 
emission  is optically thick  and decreases at higher (optically thin) 
frequencies.  The turnover occurs between 2 and 4 GHz. Their value of 
decrease at optically thin frequencies of 0.145 $\pm$ 0.005 \%~yr$^{-1}$ is 
consistent to that of the present study covering data from the 36 year period 
to 2003.

\subsection{Venus} The radio spectrum of Venus is the result of 
contributions from different depths in its atmosphere at high frequencies 
and 
from the surface at low frequencies. At the surface the temperature is 
$\approx 750$~K and the pressure $\approx 90$~bar. The millimetric emission arises 
from the higher atmosphere at h$\approx 80$~km where the effective 
temperature is 300~K. At intermediate radio frequencies the emission 
arises from the finite microwave opacity of SO$_{2}$ and H$_{2}$SO$_{4}$. 
At 33 GHz 
where the brightness temperature is $\approx 460$~K the effective altitude of 
the emission is 35~km; this is the altitude at which space probes recorded 
a temperature of 460~K and a pressure of $\approx 5$~bar. At altitudes of 
$\sim$100 km a strong Venusian diurnal effect has been measured by space 
probes. We have examined our data to determine whether this diurnal effect 
extends down to altitudes around 35 km where the 33 GHz emission arises.

\subsubsection{The spectrum of Venus} 
\begin{figure} \centering 
\mbox{\resizebox{0.48\textwidth}{!}{\includegraphics[angle=0]{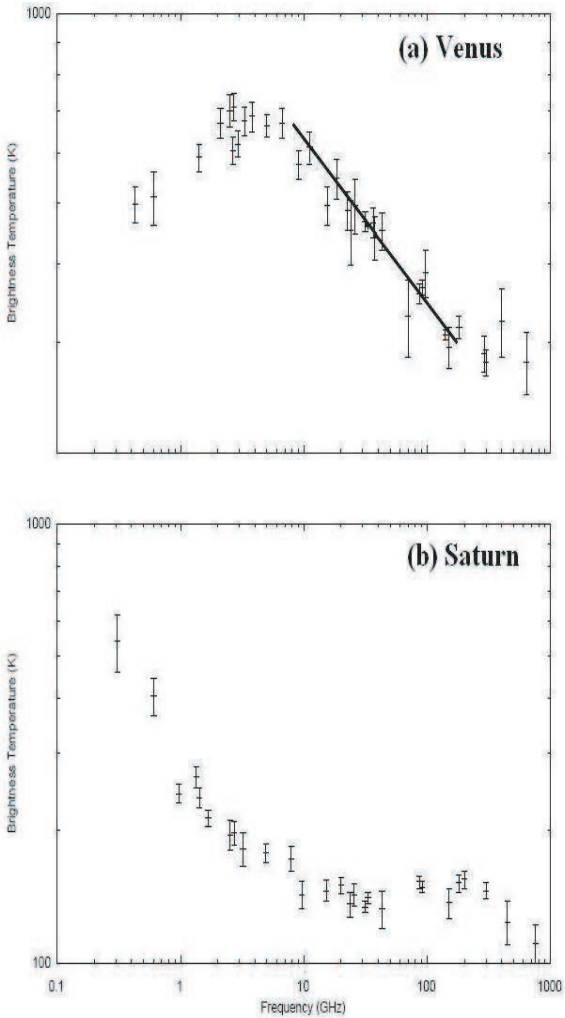}}} 
\caption{ (a) The brightness temperature spectrum of Venus from the 
literature. The full line shows the least-square best fit slope between 10 and 100~GHz 
which gives a T$_b$ spectral index at 33 GHz of $-0.278\pm0.026$.  \newline (b) 
The brightness temperature spectrum of Saturn from the literature.} 
\label{venspec} 
\end{figure}

Fig.~\ref{venspec}(a) shows the radio 
brightness temperature spectrum of Venus.
The low frequencies come from the surface region where the temperature is 
$\sim$750 K. The higher GHz frequencies arise in successively higher 
levels in the Venusian atmosphere where temperatures are lower.  The slope 
of the T$_{b}$ spectrum at 33 
GHz, estimated from a best fit to the range 10 to 100 GHz, is 
$-0.278\pm0.026$ (the temperature spectral index, $\beta$, is related to the flux density spectral index, $\alpha$, by $\beta=\alpha-2$).

\subsubsection{Brightness temperature variation with illumination} 
\begin{figure}
\centering
\mbox{\resizebox{0.48\textwidth}{!}{\includegraphics[angle=0]{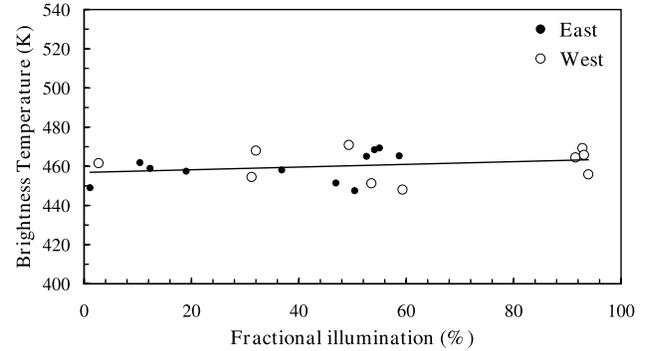}}}
\caption{ The brightness temperature of Venus plotted as function of illumination 
for the period August 2002 to September 2004. Filled circles show eastern 
illumination (surface entering darkness) and open circles show western 
illumination (leaving darkness). The full line is the best linear fit to the 
data corresponding to $+1.5\pm1.4\%$ change from zero to full illumination.} 
\label{venill} 
\end{figure}

Fig.~\ref{venill} shows the 33 GHz brightness temperature of Venus as a function of 
illumination for the period August 2002 to September 2004. During this 
period, Venus went through 1.5 Synodic cycles. Data are shown separately 
for eastern and western illumination, corresponding to times when the 
Venusian surface was entering or leaving the long period of darkness. No 
difference is seen between these two phases of illumination of more than 2 
K ( 0.3 \%). 
The 
variation of brightness temperature with illumination suggests a weak 
increase of temperature with illumination of +1.5$\pm$1.3\% (6.9$\pm$6.0 
K) from zero to full illumination.

\subsection{Saturn} Saturn is used as a secondary calibrator by the VSA. 
For accurate measurements the effect of the rings must be taken into 
account. The rings themselves emit with a brightness temperature of 10-20 
K and also block out the emission from the disk of Saturn (see for example 
Janssen \& Olsen 1978, Conway 1980; Ulich 1981). As a consequence, the 
effective brightness temperature of Saturn varies with the tilt angle 
B (the 
Saturnicentric latitude of the earth). We present data for the period 
June 2002 to August 2003. 

\subsubsection{The spectrum of Saturn} 

The 
microwave spectrum of Saturn taken from the literature is plotted in Fig.~\ref{venspec}(b) in terms of its brightness temperature. There is some uncertainty as 
to the contribution of the rings to the data presented. Since the ring 
effect is typically less than 5 percent, it lies within the error 
assigned to most data points in Fig \ref{venspec}(b). The shape of the 
spectrum is 
quantitatively similar to that of Jupiter - increasing from $\approx 130$~K 
at mm wavelengths to 540 K at 94 cm. As in the case of Jupiter there is 
evidence for the presence of ammonia as the principle source of opacity at 
radio frequencies with its inversion band at $\approx 24$~GHz. At 30 GHz the 
frequency of interest in the present study, the effective temperature 
spectral index is $\beta$ = 0.00$\pm$0.05. At frequencies of $10-43$~GHz, $\beta = -0.07\pm0.05$, while at $10-200$~GHz, $\beta = +0.041\pm0.015$. 


 \subsubsection{Secular variation of Saturn at 33 GHz} 

\begin{figure}
\centering
\mbox{\resizebox{0.48\textwidth}{!}{\includegraphics[angle=0]{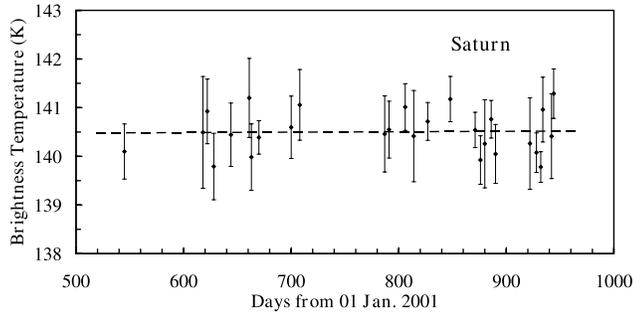}}}
\caption{ The brightness temperature of Saturn at 33 GHz for the period 29 
June 2002 to 02 August 2003. There is no secular variation greater than 
$0.4\%~yr^{-1}$ over this period.} 
\label{satshort}
\end{figure}

The VSA programme gave an accurate data set for the integrated flux 
density of Saturn extending from 29 June 2002 to 02 August 2003. During 
this period the tilt angle moved between B = 26$^\circ$ and 27$^\circ$  
with a 
quasi - annual period. The brightness temperature is then calculated by 
assuming the observed flux density arises from Saturn's disk whose 
dimensions are given in the Ephemeris. Fig \ref{satshort} is the brightness 
temperature plot over this period and shows no secular variation greater 
than 0.4\% per year at 95\% confidence level. Our results indicate that 
the mean brightness temperature of Saturn (disk plus rings) between June 
2002 and August 2003 is 140.50$\pm$0.12 K; for the data set the mean tilt 
angle B = 26.5$^\circ$.

\section{Conclusion} We have determined the flux density or brightness 
temperature at 33 GHz of 7 sources used by the VSA for calibration; the results are summarised in Tables~\ref{4}~\&~\ref{5}. This 
list includes sources widely used for calibration in this frequency range 
in radio-source and CMB studies; they are the strong sources and 
planets. The 
high level of reliability of the VSA and the stable atmosphere at the 
Iza\~{n}a site have assured accurate relative measurements of these sources. We 
use Jupiter as the reference standard in this work and adopt a brightness 
temperature of 146.6$\pm$0.75~K at 33 GHz as determined in the WMAP survey (Hill et al.~2008). 
Any refinement in the future can then be applied to the present results. 
In 
order to make our results useful for accurate flux density/brightness 
temperature measurements in the frequency band around 33 GHz we derive a 
spectral index 
from published data.

\begin{table*} \centering \caption{Secular change in flux density (or 
brightness temperature) of calibrators relative to Jupiter at 33 GHz 
(epoch 2001.0).}
\label{4} 
\begin{tabular}{cc} \hline \noalign{\smallskip} 
Radio sources &dS/Sdt (\%~yr$^{-1}$) \\
\hline\noalign{\smallskip}
Cas~A &$-0.394\pm0.019$\\
Tau~A &$-0.22\pm0.07$\\
NGC7027$^{1}$ &$-0.16\pm0.04$\\
Cyg~A &$+0.043\pm0.039$\\
Venus$^{2}$ &$+1.5\pm1.3$\\
\hline
\multicolumn{2}{l}{1 At optically thin frequencies; see Section 4.4.2.}\\
\multicolumn{2}{l}{2 Secular variation of T$_{b}$ with illumination; see 
Section 4.5.2.}\\
\end{tabular}
\end{table*}

Since we have observations of the calibrators 
extending over nearly 4 years, we are able to make an accurate assessment 
of the calibrator variation over this period. With rms individual 
comparison measurements of 0.4 to 1.0 \% depending on the time separation 
of the pair of observations, we are sensitive to small secular flux 
density or brightness temperature variations. A significant secular decrease in flux density was 
established 
for Cas~A and Tau~A over the 4-year observing period. By combining our 
data with historical data in the literature, we have also demonstrated a 
significant secular decrease for NGC7027 at optically thin frequencies. 
Over this period the secular 
variation of the two sources expected not to vary, Cyg~A and Jupiter, was 
less than 0.1\%~yr$^{-1}$. Venus showed a slight (1.5$\pm$1.3\%) variation with illumination.


\begin{table*} \centering \caption{Summary of 
calibrator flux densities or brightness temperatures at 33 GHz and epoch 
2001.0, all relative to an adopted Jupiter T$_{b}$ = 146.6~K.} 
\label{5} 
\begin{tabular}{cccc} \hline\noalign{\smallskip} 
Radio sources &Flux density &Error &Spectral index\\
 &(Jy) &(Jy) &(at 33 GHz)\\
\hline\noalign{\smallskip}
Cas~A &182.0 &0.1 &$-0.69\pm0.02$\\
Cyg~A &36.4 & 0.2 &$-1.208\pm0.017$\\
Tau~A$^{1,2}$ &322 &4 &$-0.32\pm0.01$\\
NGC7027$^3$ &5.39 &0.04 &$-0.119\pm0.025$\\
\\
\hline\noalign{\smallskip}
Planets &Brightness temperature &Error &Spectral index\\
 &(K) &(K) &(at 33 GHz)\\
\hline\noalign{\smallskip}
Jupiter &146.6 &0.75 &$+0.248\pm0.004$\\
Venus &460.3 &3.2 &$-0.278\pm0.026$\\
Saturn$^4$ &140.50 &0.12 &$0.00\pm0.05$\\
\noalign{\smallskip}\hline
\multicolumn{4}{l}{$^1$Tau~A is 7.8$\pm$0.6\%~yr$^{-1}$ linearly polarized 
at pa $= 148^\circ \pm3^\circ$.}\\
\multicolumn{4}{l}{$^2$Tau~A flux density is 332.8$\pm$0.7 Jy at pa = 
0$^\circ$.}\\
\multicolumn{4}{l}{$^3$Flux density at 2003.0.}\\
\multicolumn{4}{l}{$^4$Flux density at 2003.0; tilt angle B of rings is 
26.5$^\circ$.}\\
\end{tabular}
\end{table*}

\subsection*{Acknowledgements}
YH thanks the King Abdul City of Science and Technology for support. CD acknowledges support from the U.S. {\it Planck} project, which is funded by the NASA Science Mission Directorate.

\bibliographystyle{mnras}

\end{document}